# Structured Spreadsheet Modelling and Implementation with Multiple Dimensions - Part 2: Implementation

*Paul Mireault*

*Founder, SSMI International*
*Honorary Professor, HEC Montréal*
*Paul.Mireault@SSMI.International*

**ABSTRACT**

*In Part 1, we showed how to develop a conceptual model of a problem involving variables of multiple dimensions, like Products, Regions, Sectors and Months. The conceptual model is presented as a Formula Diagram, giving a global view of the interaction between all the variables, and a Formula List, giving a precise view of the interaction between the variables. In this paper, we present precise steps to implement a multi-dimensional problem in a way that will produce a spreadsheet that is easy to maintain.*

## 1   Introduction

Dimensions are an integral part of many models we use every day. Without thinking about it, we frequently use the time dimension: many financial and accounting spreadsheets have columns representing months or years. Representing a second dimension is often done by repeating blocs of formulas in a worksheet or creating multiple worksheets with the same structure. Adding a third or a fourth dimension is a perilous operation, resulting in structures that are hard to understand and maintain.

*Figure 1 United Fruits 3-Dimension Spreadsheet*

*Figure 23-Dimension example from the Microsoft Help on the INDEX Function*

Problems with the same dimensions are implemented with different structures depending on the developer's preferences. Figure 1, from Brandewindere (2018), shows an implementation with three dimensions: *Product*, *Location* and *Quarter*. The developer decided to implement the products in different worksheets, the quarters in columns and the locations in repeated blocks where we can see three variables. Figure 2, from the example on the use of the Index function, Microsoft (2018), shows an implementation where the developer decided to implement one variable in two-dimensional tables for the *Location* and *Quarter* dimensions, replicated for each *Product*.

Past research has studied the problem of multiple dimensions in spreadsheet. Cunha, Fernandes, Mendes, Pacheco, and Saraiva (2012)and Cunha, Erwig, Mendes, and Saraiva (2016) have shown how to create an entity-relationship model from an existing spreadsheet.

Other research has focused on building a conceptual model before implementing the spreadsheet. Rajalingham, Chadwick, and Knight (2001)proposed graph-based conceptual model and Mireault (2017c) described a diagram-based conceptual model and introduced a representation of 1-dimensional variables.

In Part 1, Mireault (2017b), we described how to develop a conceptual model of a multidimensional problem using a small case, Acme Techno Widgets, which is reproduced in the Appendix. The conceptual model is composed of a Formula Diagram and a Formula List. In the next section, we describe the important concepts of data warehouse design that we use in the Excel implementation of the conceptual model. In Section 3, we will then describe the steps needed to create the structure spreadsheet. Then, in Section 4, we will illustrate the work necessary to modify the spreadsheet when we add a new element to a dimension. Finally, we will conclude with a discussion of the work involved in implementing multi-dimensional spreadsheets.

## 2 Data Warehouse Design and Implementation Concepts

We base our structured implementation on principles used in designing data warehouses. One common data structure is called the star schema. A star represents a multidimensional table, called a fact table, linked to other tables, called dimension tables. A dimension table contains a primary key, which identifies a precise row of table, and characteristics, which describe the dimension element itself. A constellation schema is an extension of the star schema that has multiple fact tables sharing dimension tables (Vaisman and Zimányi (2014).)

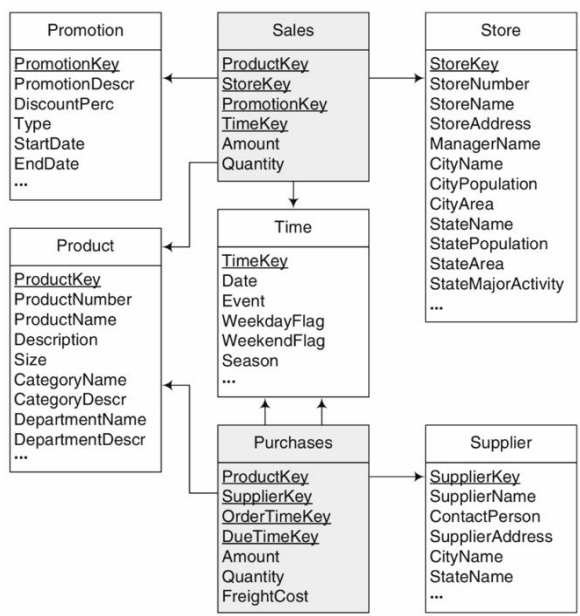

*Figure 3 Constellation Schema (from Vaisman and Zimányi (2014), page 125)*

In Figure 3, we see that the primary keys are underlined. The Sales table has a primary key composed of ProductKey, StoreKey, PromotionKey and TimeKey. Each of the components serves as a foreign key (in the parent table Sales) to refer to the primary keys of children tables Product, Store, Promotion and Time. Thus, even though the Sales table only contains an Amount and a Quantity, the foreign keys let us associate them to a specific store, product, promotion and moment in time.

In the SSMI implementation, we will consider each repeating group as either a dimension table, a fact table, or both. It will be considered as a fact if it has at least one calculated variable. Base repeating groups, those having a dimension set of size 1, will correspond to dimension tables.

Fact tables have a primary key and one foreign key for each dimension table they are joined to. In our implementation, we will use the same concepts, as described in Mireault (2017a).

We can reference the parent of a foreign key with an `INDEX-MATCH` formula:

```
INDEX(Value, MATCH(Foreign Key, Primary Key, exact match code))
```

where `Primary Key` and `Value` are in the parent worksheet, and `Foreign Key` is in the child worksheet.

## 3   Spreadsheet Implementation
### 3.1   Naming Convention

In databases, we often see that the foreign key in a child table has the same name as the primary key of the parent table it is linked to. There is no confusion because a name is associated with the table it belongs to. Thus, the two names `CLIENT_ID` are in fact `ORDER.CLIENT_ID` and `CLIENT.CLIENT_ID`, and the full names must be used whenever the two tables are used in the same operation.

In Excel, names can either have a global scope, called Workbook, or a local scope, associated to a specific worksheet, as shown in Figure 4(a). When we use the Create from selection button, Excel always creates a name with the Workbook scope if there isn't already one, in which case it will create it in the worksheet's scope.

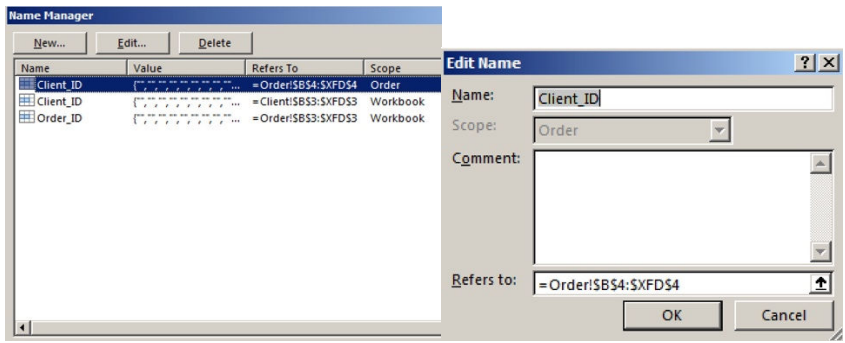

(a) (b)

*Figure 4 Excel Names with Scopes*

Since Excel does not provide a way to change the scope of a name after its creation, as shown by the grey field in Figure 4(b), we decided to always use global names in an SSMI implementation. This means that we need to avoid using the same name more than once. Therefore, we use the following naming convention for primary keys and foreign keys:

- A primary key is defined in its principal data worksheet, like **Client ID** in the *ClientData* worksheet and **Order ID** in the *OrderData* worksheet.

- A foreign key starts with the name of the primary key it refers to, followed by "in" and the name of its worksheet, like **Client_ID_in_Order** in the *OrderData* worksheet.

By using this naming convention, all reference formulas have the following form:

```
INDEX(Variable, MATCH(Foreign Key, Primary Key, Exact match code))
```

**Variable** and **Primary Key** are in the same dimension set. **Foreign Key** is in the current dimension set, i.e. the one in which the INDEX reference function is located. Figure 5 shows an example where we extract the name of the client of order 25.

(a) Child worksheet                    (b) Parent worksheet

*Figure 5 Using a Foreign Key to get data from a Parent*

Aggregate functions can use the family of IF and IFS functions provided in Excel: AVERAGEIF, AVERAGEIFS, COUNTIF, COUNTIFS, MAXIFS, MINIFS, SUMIF and SUMIFS.

### 3.2   Model Management Variables

To reduce the possibility of having an incorrect number of columns, we can use model management variables.

1. For each dimension set, we can calculate the number of columns it needs. We can also calculate the coordinate of the last column, taking into account the variable labels in column A and any needed initialization columns.

2. For each worksheet, we can verify that it has the correct number of columns with a count of its primary key.

3. For each variable from a different dimension set, since we usually reference its primary key and its foreign keys,

In the structured implementation methodology, we recommend copying whole columns, encompassing all the variables at the same time, we can perform validation 2 once per worksheet. Since the use of aggregate formulas is rarer, we would then perform validation 3 each time it is used.

### 3.3 Structured Implementation

The steps regarding the structured implementation have been developed to facilitate the model's implementation and its maintenance. The basic rule is that a worksheet only contains the definition formulas of variables belonging to the same dimension set. This makes it easier to make sure that we have the correct number of columns in each worksheet.

This doesn't mean that all the variables used in the worksheet have the same number of columns: variables that are used in an aggregate formula are usually from a dimension set with more columns.

#### 3.3.1 Determine the needed worksheets

From the Formula Diagram, we first determine the data sheets by listing all the dimension sets that have at least one data variable or input variable. Dimensionless data and input variables appear in the *Data* worksheet. The other variables are in worksheets named by their dimension set followed by the word *Data*.

*Table 1 The Data worksheets*

| Worksheet | Data or Input Variable |
| --- | --- |
| *Data* | **Base Price** and **Monthly Fixed Cost** |
| *ProductData* | **Base Price Multiplier** and **Unit Production Cost** |
| *SectorData* | **Rebate Percentage**, **DemPar A** and **DemPar B**. |
| *RegionData* | **Unit Delivery Cost** |
| *Sector-Product Data* | **Product Distribution per Sector** |
| *Sector-RegionData* | **Region Sales Distribution per Sector** |
| *Month-SectorData* | **Monthly Sales Distribution per Sector**. |

Then, we determine the model sheets from the dimension sets that have at least one calculated or output variable. Dimensionless calculated and output variables are defined in the *Model* worksheet. The other variables are defined in worksheets named by their dimension set.

*Table 2 The Model worksheets*

| Worksheet | Calculated or Output Variable |
| --- | --- |
| *Sector* | **Sector Price Factor**, **Sector Base price** and **Sector Annual Demand Units**. |
| *Product-Region* | **PR Unit Cost** |
| *Sector-Product* | **Annual Sector-Product Unit Sales**, and **Annual Sector-Product Sales Amount** |
| *Month-Sector-Product* | **MSP Unit Sales** and **MSP Sales Amount** |
| *Month-Sector-Product-Region,* shortened to *MSPR* | **MSPR Unit Sales** and **MSPR Variable Cost** |
| *Month* | **Monthly Variable Cost**, **Monthly Unit Sales**, **Monthly Sales Amount**, **Monthly Costs** and **Monthly Profit** |
| *Month-Product-Region* | **MPR Unit Sales** |
| *Month-Product* | **MP Unit Sales** and **MP Sales Amount** |
| *Model* | **Total Profit** |

#### 3.3.2 Determine the Primary keys and the Foreign keys

To avoid having to manipulate names that are very long, we will use a dimension's initial when we create primary and foreign key names. Thus, we will use *PR in MSPR* to refer to *Product-Region in Month-Sector-Product-Region* and *SP in MSP* to refer to *Sector-Product in Month-Sector-Project*.

Every dimension set used needs a primary key: **Product**, **Sector**, **Region**, **Month**, Sector-Product (**SP**), Month-Sector (**MS**), Sector-Region (**SR**), Month-Product (**MP**), Product-Region (**PR**), Month-Sector-Product (**MSP**), Month-Product-Region (**MPR**) and Month-Sector-Product-Region (**MSPR**).

Every time a formula uses a variable from another dimension-set, we need a foreign key. There are two cases: aggregate and non-aggregate formulas.

An aggregate formula, like `Monthly Unit Sales = SUM(MSPR Unit Sales)`, has a resulting dimension set, *Month*, and a starting dimension set, *Month-Sector-Product-Region*. We will create a Foreign Key with the form *Resulting Dimension set in starting Dimension set*. (As explained in Part 1, the *Resulting Dimension Set* is a subset of the *Starting Dimension Set*.) Table 3 lists the foreign keys created in this step.

*Table 3 Foreign Keys created from aggregate formulas*

| Var No | Foreign Key |
|---|---|
| 22 | *M in MSPR* |
| 23 | (Same as 22) |
| 24 | *M in MSP* |
| 28 | *MPR in MSPR* |
| 29 | *MP in MSP* |
| 30 | (same as 29) |

*Table 4 Foreign Keys created from non-aggregate formulas*

| Var No | Foreign Key (variable requiring the foreign key) |
|---|---|
| 11 | *P in PR* (**Unit Production Cost**) and *R in PR* (**Unit Delivery Cost**) |
| 13 | *S in SP* (**Sector Annual Demand Units**) |
| 14 | *S in SP* (**Sector Base Price**) and *P in SP* (**Base Price Multiplier**) |
| 18 | *SP in MSP* (**Annual Sector Product Unit Sales**) and *MS in MSP* (**Monthly Sales Distribution per Sector**) |
| 19 | (same as 18) |
| 20 | *MSP in MSPR* (**MSP Unit Sales**) and *SR in MSPR* (**Region Sales Distribution per Sector**) |
| 21 | *PR in MSPR* (**PR Unit Cost**) |

In a non-aggregate formula, like `Price = Sector Base Price * Base Price Multiplier`, the resulting dimension set, *Sector-Product* is the union of the dimension sets of its components, *Sector* for **Sector Base Price** and *Product* for **Base Price Multiplier**. We will create the foreign keys with the form *Starting Dimension Set in Resulting Dimension Set*. Table 4 lists the Foreign Keys created.

Figure 6 shows the Primary Key for dimension set *Month-Sector-Product*, **MSP**, and all its possible Foreign Keys. In our example, we only need Foreign Keys **MS in MSP** and **SP in MSP**.

| | A | B | C | D | E | F | G | H | I | J | K |
|---|---|---|---|---|---|---|---|---|---|---|---|
| 1 | MSP | | Jan-G-S | Jan-G-D | Jan-M-S | Jan-M-D | Jan-P-S | Jan-P-D | Jan-E-S | Jan-E-D | Feb-G-S |
| 2 | M in MSP | | Jan | Jan | Jan | Jan | Jan | Jan | Jan | Jan | Feb |
| 3 | S in MSP | | G | G | M | M | P | P | E | E | G |
| 4 | P in MSP | | S | D | S | D | S | D | S | D | S |
| 5 | MS in MSP | | Jan-G | Jan-G | Jan-M | Jan-M | Jan-P | Jan-P | Jan-E | Jan-E | Feb-G |
| 6 | MP in MSP | | Jan-S | Jan-D | Jan-S | Jan-D | Jan-S | Jan-D | Jan-S | Jan-D | Feb-S |
| 7 | SP in MSP | | G-S | G-D | M-S | M-D | P-S | P-D | E-S | E-D | G-S |

*Figure 6 Primary Key MSP and all its possible Foreign Keys*

Constructing the Primary Keys and Foreign Keys can be as simple as importing them from a database. They can also be constructed using the technique presented in Mireault (2016).

### 3.3.3 The Model Worksheets

Implementing the formulas from the Formula List is straightforward. Following the SMMI implementation methodology, every variable's definition formula is presented in a block with the top part consisting of references to the variables used in the definition and the bottom part is the formula using the cells just above.

We need to consider the following cases:

- Non-aggregate formula.
  - All variables are from the resulting variable's dimension set. This is the case of calculating **Annual Sector-Product Sales Amount = Annual Sector-Product Unit Sales*Price**.
  - All variables are from a subset of the resulting variable's dimension set. This is the case of **Price = Sector Base Price + Base Price Multiplier**.
  - There can be a mix of the two, with some variables from the same dimension set as the result and others from a subset. This is the case of **Annual Sector-Product Unit Sales = Sector Annual Demand Units * Product Distribution per Sector**.
- Aggregate formula. This is the case of **MP Unit Sales = SUM(MSP Unit Sales)**.

The three examples of non-aggregate formulas presented above come from the *Sector-Product* worksheet, illustrated in Figure 7.

|   | A | B | C | D | E | F | G | H | I | J |
|---|---|---|---|---|---|---|---|---|---|---|
| 1 | Sector-Product | J1 | | | | | | | | |
| 2 | | | | | | | | | | |
| 3 | SP | | G-S | G-D | M-S | M-D | P-S | P-D | E-S | E-D |
| 4 | | | | | | | | | | |
| 5 | Sector Annual Demand Units | | 2718.07 | 2718.07 | 1787.02 | 1787.02 | 4605.16 | 4605.16 | 4686.71 | 4686.71 |
| 6 | Product Distribution per Sector | | 65% | 35% | 25% | 75% | 40% | 60% | 80% | 20% |
| 7 | Annual Sector-Product Unit Sales | | 1766.74 | 951.32 | 446.75 | 1340.26 | 1842.06 | 2763.09 | 3749.37 | 937.34 |
| 8 | | | | | | | | | | |
| 9 | Sector Base Price | | $84.00 | $84.00 | $112.00 | $112.00 | $126.00 | $126.00 | $42.00 | $42.00 |
| 10 | Base Price Multiplier | | 1 | 1.45 | 1 | 1.45 | 1 | 1.45 | 1 | 1.45 |
| 11 | Price | | $84.00 | $121.80 | $112.00 | $162.40 | $126.00 | $182.70 | $42.00 | $60.90 |
| 12 | | | | | | | | | | |
| 13 | Annual Sector-Product Unit Sales | | 1766.74 | 951.32 | 446.75 | 1340.26 | 1842.06 | 2763.09 | 3749.37 | 937.34 |
| 14 | Price | | $84.00 | $121.80 | $112.00 | $162.40 | $126.00 | $182.70 | $42.00 | $60.90 |
| 15 | Annual Sector-Product Sales Amount | | $148,406 | $115,871 | $50,036 | $217,658 | $232,100 | $504,817 | $157,473 | $57,084 |

*Figure 7 The Sector-Product model worksheet*

In the case of non-aggregate formulas, the reference formula to a variable of the same dimension set is simply the name of the variable, as illustrated in rows 6, 13 and 14 of Figure 8. The reference formulas of variables defined in a subset of the current dimension set, **Sector Annual Demand** and **Sector Base Price** in *Sector* and **Base Price Multiplier** in *Product*, use the `INDEX–MATCH` form presented in section 3.1. This is illustrated in rows 5, 9 and 10 of Figure 8.

|   | A | B | C |
|---|---|---|---|
| 1 | Sector-Product | =Last_SP_column | |
| 2 | | | |
| 3 | SP | | =SP |
| 4 | | | |
| 5 | Sector Annual Demand Units | | =INDEX(Sector_Annual_Demand_Units,MATCH(S_in_SP,Sector_Code,0)) |
| 6 | Product Distribution per Sector | | =Product_Distribution_per_Sector |
| 7 | Annual Sector-Product Unit Sales | | =C5*C6 |
| 8 | | | |
| 9 | Sector Base Price | | =INDEX(Sector_Base_Price,MATCH(S_in_SP,Sector_Code,0)) |
| 10 | Base Price Multiplier | | =INDEX(Base_Price_Multiplier,MATCH(P_in_SP,Product_Code,0)) |
| 11 | Price | | =C9*C10 |
| 12 | | | |
| 13 | Annual Sector-Product Unit Sales | | =Annual_Sector_Product_Unit_Sales |
| 14 | Price | | =Price |
| 15 | Annual Sector-Product Sales Amount | | =C13*C14 |

*Figure 8 The Formula View of the Sector-Product model worksheet*

Implementing an aggregate calculation follows a similar block structure, as shown in Figure 9. Row 6 is not used in the calculation: it only serves in documenting the dimension set of the variable that is being aggregated, **MSP Unit Sales** in this case. The coloured highlights show the mechanics of the SUMIF formula being used to perform the calculation: we show the cells in used in the criteria in green and the cells used in the calculation in yellow.

|    | A | B | C | D | E | F | G | H | I | J | K | L |
|----|---|---|---|---|---|---|---|---|---|---|---|---|
| 1  | Month-Product | Z1 | | | | | | | | | | |
| 2  | | | | | | | | | | | | |
| 3  | MP | | Jan-S | Jan-D | Feb-S | Feb-D | Mar-S | Mar-D | Apr-S | Apr-D | May-S | May-D |
| 4  | | | | | | | | | | | | |
| 5  | Last MSP column | CT1 | | | | | | | | | | |
| 6  | MSP | | Jan-G-S | Jan-G-D | Jan-M-S | Jan-M-D | Jan-P-S | Jan-P-D | Jan-E-S | Jan-E-D | Feb-G-S | Feb-G-D |
| 7  | MSP Unit Sales | | 159.01 | 85.62 | 35.74 | 107.22 | 221.05 | 331.57 | 224.96 | 56.24 | 176.67 | 95.13 |
| 8  | MP in MSP | | Jan-S | Jan-D | Jan-S | Jan-D | Jan-S | Jan-D | Jan-S | Jan-D | Feb-S | Feb-D |
| 9  | MP | | Jan-S | Jan-D | Feb-S | Feb-D | Mar-S | Mar-D | Apr-S | Apr-D | May-S | May-D |
| 10 | MP Unit Sales | | 640.76 | 580.65 | 719.46 | 594.68 | 759.91 | 581.22 | 769.50 | 562.14 | 812.87 | 557.15 |

*Figure 9 Aggregate calculation of MP Unit Sales*

Figure 10 shows the formula view of the calculation. In each column of row 10, the SUMIF function scans row 8 looking for values that are equal to the value of row 9 above and adds the values of row 7 when it finds them.

|    | A | B | C | D | E | F | G |
|----|---|---|---|---|---|---|---|
| 1  | Month-Product | =Last_MP_column | | | | | |
| 2  | | | | | | | |
| 3  | MP | | =MP | =MP | =MP | =MP | =MP |
| 4  | | | | | | | |
| 5  | Last MSP column | =Last_MSP_column | | | | | |
| 6  | MSP | | =MSP | =MSP | =MSP | =MSP | =MSP |
| 7  | MSP Unit Sales | | =MSP_Unit_Sales | =MSP_Unit_Sales | =MSP_Unit_Sales | =MSP_Unit_Sales | =MSP_Unit_Sales |
| 8  | MP in MSP | | =MP_in_MSP | =MP_in_MSP | =MP_in_MSP | =MP_in_MSP | =MP_in_MSP |
| 9  | MP | | =MP | =MP | =MP | =MP | =MP |
| 10 | MP Unit Sales | | =SUMIF(8:8,9:9,7:7) | =SUMIF(8:8,9:9,7:7) | =SUMIF(8:8,9:9,7:7) | =SUMIF(8:8,9:9,7:7) | =SUMIF(8:8,9:9,7:7) |

*Figure 10 Formula view of the aggregate calculation of MP Unit Sales*

Figure 9 also illustrates the use of model management formulas discussed in section 3.2. All the variables belonging to dimension set *Month-Product* extend to column Z and all those belonging to *Month-Sector-Product* to column CT.

### 3.3.4 Interface Worksheet

The Interface worksheet is where the users will interact with the spreadsheet model. It is where they will enter the value of the Input variables and observe the resulting values of the Output variables. This is where spreadsheet developers can prepare dashboards, tables and charts.

Presenting the values of dimensionless or 1-dimension variables, like **Total Profit** and **Total Monthly Sales**, is straightforward: they can simply be reference formulas.

We need an area to prepare the codes that will be used in one `INDEX-MATCH` references, in the same structure that will be used the presentation.

With one dimension, it's straightforward.

With Two-dimensions, we need to decide which dimension will be in rows and which one will be in columns.

With Three dimensions, we have no choice but to present them in blocks. We need to decide which dimension will be repeated in blocks: the one with the smallest number of values is usually a good choice because it reduces the number of blocks that need to be repeated.

To prepare the primary keys, we set up the base codes in the visual structure we want, and we build the primary key of the dimension set by concatenating the values of the dimension codes, as illustrated in Figure 11(a). In the presentation area, we use a reference formula using the relative coordinate of the primary key we built in the preparation worksheet: = `INDEX(value, MATCH(presentation code, Primary key of Dimension set, Exact match code)`, as shown in Figure 11(b).

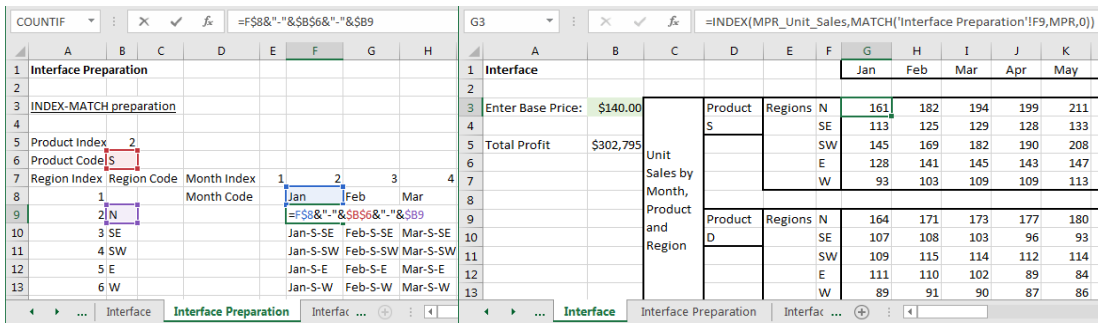

(a) Preparing the primary key     (b) Using the primary key

*Figure 11 Preparing and using a primary key for the Interface sheet to display the MPR Unit Sales output variable*

## 4    Maintaining a structured implementation

In this section, we present how to use the Formula Diagram to determine the exact operations needed to change the number of instances in a structured implementation.

When you change the number of members (instances) in a basic dimension, you know exactly which worksheets to modify.

- The basic dimension worksheet
- All the worksheet with a dimension set containing the base dimension
- All the worksheet variables with an aggregate calculation using a variable from a worksheet above. They are easy to recognize in the Formula Diagram as variables calculated with an arrow coming from repeating sub-model with the base dimension

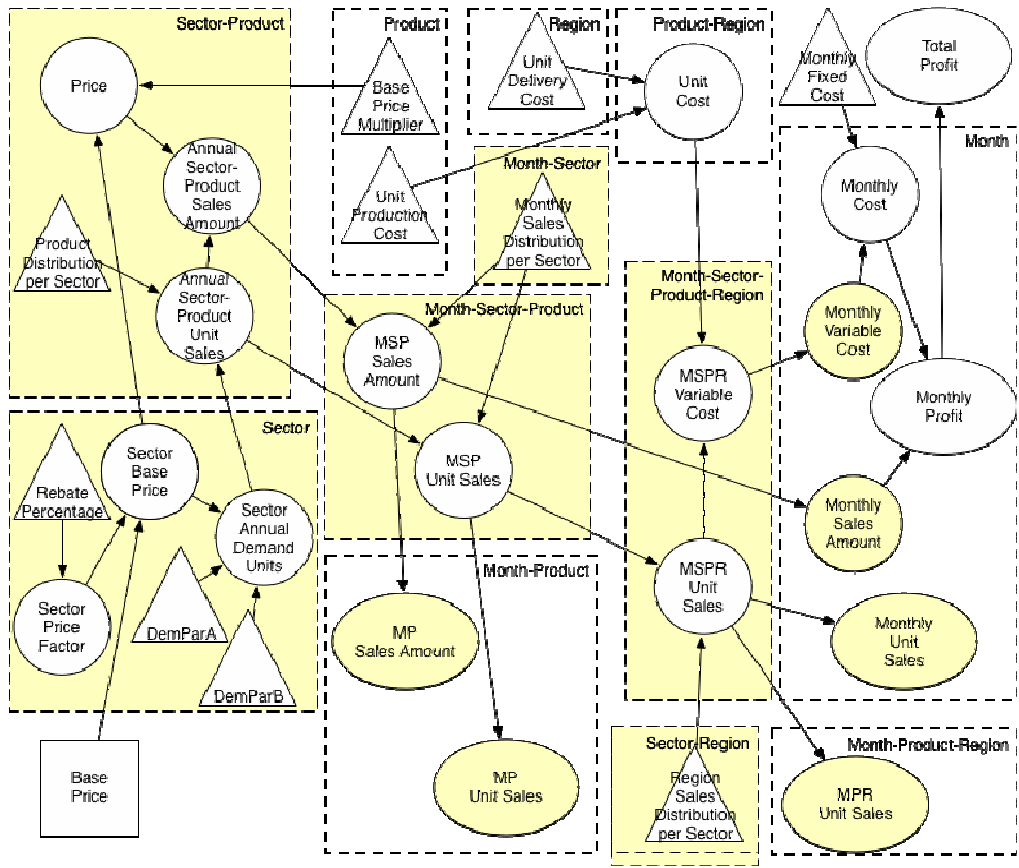

*Figure 12 Impact of adding a Sector*

In our example adding a sector requires the following operations, as illustrated in Figure 12:

- Add one new column in sheets *Sector* and *Sector-Data*
- Add the appropriate number of columns in *Sector-Products*, *Month-Sector*, *Month-Sector-Product*, *Sector-Region*, *Month-Sector-Product-Region* and their corresponding *Data* worksheets. (The number columns to add is the product of the cardinalities of the other dimensions. For the dimension set *Month-Sector-Product* it is: Number of Months × Number of Products. We could also use model management variables that calculate the last column of each dimension set, as explained in section 3.2 above.)
- Add columns to the aggregation formulas of variables **MP Sales Amount**, **MP Unit Sales**, **MPR Unit Sales**, **Monthly Variable Cost**, **Monthly Sales Amount** and **Monthly Unit Sales**. There is no need to change the formulas.

# 5  Conclusion

In this paper, we use proven concepts from data warehouse design and software engineering to implement spreadsheets that are hard to implement.

As shown in the Introduction, developers tend to develop their spreadsheet according to the way they want to present it to the user. We use the three-tier architecture, from software engineering, to develop the calculations separately from their presentation.

Savage (1997) described two important problems with using dimensions in spreadsheets. First is *scalability*, which involves changing the cardinality of a dimension. He concluded that spreadsheets rarely scale well. Second is *hyper-scalability*, which involves changing the dimensions themselves, such as adding more dimensions. His conclusion was, succinctly, "Forget it".

By using a structure similar to well established data warehouse design concepts, we showed, in section 4, that scalability can be well managed. We also showed, in section 3.3, that hyper-scalability is possible when we use a conceptual model to guide us through the implementation.

**Appendix – Case Study**

In this section, we present a pedagogical case study to illustrate the concepts presented in this paper. Some numbers are changed from Part 1, but that has no impact on the model.

# Acme Techno Widgets Company

The Acme Techno Widgets Company (ATW) produces and sells widgets. Its sales force is assigned to four major sectors: Government, military, education and private. It produces two products, the Standard widget and the Deluxe widget.

Market research has established that the annual demand for widgets depends on each sector's Standard widget price. The Pricing Director explains:

*We start by setting a global base price. Then, for each sector, we tell our sales force that they can offer a rebate. For instance, we offer a 70% rebate to the education sector and it's 10% for the private sector because purchases are usually made by researchers with limited funds. The military sector gets a 20% rebate and the government 40%. This is not made public: all our price lists show the base price, but our clients in each sector are aware of the rebate they can get.*

*Each sector reacts differently to a change of price. We consulted with a market research expert and she came up with multiple demand functions, one for each sector. The demand function estimates a sector's annual demand for a given base price. The demand function has the form $B/Price^A$. The parameters $A$ and $B$ are different for each sector, and $Price$ is the sector's price, after the rebate. This table shows the values the expert gave us:*

| Sector | Government | Military | Private Sector | Education |
|---|---|---|---|---|
| **Rebate Percentage** | 40% | 20% | 10% | 70% |
| **DemParA** | 3.59 | 3.46 | 3.18 | 4.11 |
| **DemParB** | 22000000000 | 22000000000 | 22000000000 | 22000000000 |

*The price of the Deluxe widget is 45% higher than the Standard widget.*

The Sales Manager explains the sales pattern:

*The annual demand of each Sector is split between the Standard and Deluxe products, but the distribution is very different in each sector. For instance, in the education sector, with its limited funds, the split is 80%-20% and it is 25%-75% in the military sector. I guess these guys always go for the best, and they have higher budgets. The distribution is 65%-35% for the government sector and 40%-60% for the private sector. The ratios are then applied to the sector's annual demand to get the annual demand by product.*

*Another interesting pattern is the distribution of sales during the year. We noticed that our clients buy more just before the end of their fiscal year, when some want to spend their budget surpluses, and the beginning, when others have new funds allotted. Each sector has a different pattern, and we noticed that it is pretty stable year after year.*

|  | Government | Military | Private Sector | Education |
|---|---|---|---|---|
| Jan | 9% | 8% | 12% | 6% |
| Feb | 10% | 9% | 11% | 8% |
| Mar | 12% | 10% | 9% | 9% |
| Apr | 12% | 12% | 7% | 10% |
| May | 11% | 13% | 6% | 12% |
| Jun | 9% | 11% | 4% | 12% |
| Jul | 7% | 9% | 5% | 11% |
| Aug | 6% | 7% | 6% | 9% |
| Sep | 5% | 6% | 8% | 7% |
| Oct | 5% | 4% | 9% | 6% |
| Nov | 6% | 5% | 11% | 5% |
| Dec | 8% | 6% | 12% | 5% |
| Total | 100% | 100% | 100% | 100% |

*Sales to a sector are not uniformly distributed by region. For example, there are more universities in the South-West than in the West. The following table shows the distribution of a sector's sales by region. With it, we can calculate the expected monthly sales per product per region, which helps our Logistics Department do its planning.*

| | Government | Military | Private Sector | Education |
|---|---|---|---|---|
| N | 25% | 52% | 22% | 24% |
| SE | 18% | 13% | 21% | 15% |
| SW | 18% | 18% | 17% | 32% |
| E | 22% | 0% | 25% | 17% |
| W | 17% | 17% | 15% | 12% |
| Total | 100% | 100% | 100% | 100% |

The costs of producing a widget are $48 and $72 for the Standard and the Deluxe widget respectively. The monthly fixed costs for this year are $20000. Delivery costs depend solely on the region and are shown in this table:

| Region | North | South-East | South-West | East | West |
|---|---|---|---|---|---|
| **Unit Delivery Cost** | $10.25 | $9.73 | $9.58 | $8.26 | $11.02 |

The company CEO wants to see the following results:
- The monthly unit sales per product per region.
- The monthly sales amount and unit sales per product.
- The monthly unit sales and profit.
- The total profit.

## Acme Techno Widget Company Formula Diagram

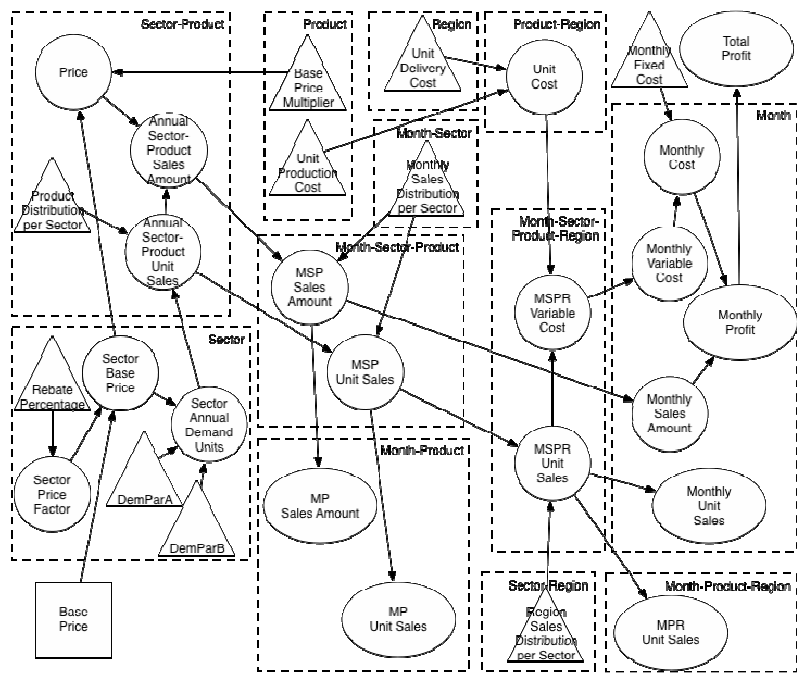

## Acme Techno Widget Company Formula List

| Var No | Variable | Type | Dimension Set | Value / Formula |
|---|---|---|---|---|
| 1 | Base Price | Input | | $100 |
| 2 | Base Price Multiplier | Data | Product | (1, 1.45) |
| 3 | Unit Production Cost | Data | Product | list of values |
| 4 | Rebate Percentage | Data | Sector | list of values |
| 5 | Sector Price Factor | Calculated | Sector | 1-Rebate Percentage |
| 6 | Sector Base Price | Calculated | Sector | Base Price * Price Factor |
| 7 | DemParA | Data | Sector | list of values |
| 8 | DemParB | Data | Sector | list of values |
| 9 | Sector Annual Demand Units | Calculated | Sector | DemParA*DemParB^-Sector Base Price |
| 10 | Unit Delivery Cost | Data | Region | list of values |
| 11 | PR Unit Cost | Calculated | Product-Region | Unit Production Cost + Unit Delivery Cost |
| 12 | Product Distribution per Sector | Data | Sector-Product | list of values |
| 13 | Annual Sector-Product Unit Sales | Calculated | Sector-Product | Sector Annual Demand Units * Product Distribution per Sector |
| 14 | Price | Calculated | Sector-Product | Sector Base Price * Base Price Multiplier |
| 15 | Annual Sector-Product Sales Amount | Calculated | Sector-Product | Annual Sector-Product Unit Sales * Price |
| 16 | Region Sales Distribution per Sector | Data | Sector-Region | list of values |
| 17 | Monthly Sales Distribution per Sector | Data | Month-Sector | list of values |
| 18 | MSP Unit Sales | Calculated | Month-Sector-Product | Annual Sector-Product Unit Sales * Monthly Sales Distribution per Sector |
| 19 | MSP Sales Amount | Calculated | Month-Sector-Product | Annual Sector-Product Sales Amount * Monthly Sales Distribution per Sector |
| 20 | MSPR Unit Sales | Calculated | Month-Sector-Product-Region | MSP Unit Sales * Region Sales Distribution per Sector |
| 21 | MSPR Variable Cost | Calculated | Month-Sector-Product-Region | MSPR Unit Sales * PR Unit Cost |
| 22 | Monthly Variable Cost | Calculated | Month | SUM(MSPR Variable Cost) |
| 23 | Monthly Unit Sales | Output | Month | SUM(MSPR Unit Sales) |
| 24 | Monthly Sales Amount | Calculated | Month | SUM(MSP Sales Amount) |
| 25 | Monthly Fixed Cost | Data | | $20000 |
| 26 | Monthly Costs | Calculated | Month | Monthly Fixed Cost + Monthly Variable Cost |
| 27 | Monthly Profit | Calculated | Month | Monthly Sales Amount - Monthly Costs |
| 28 | MPR Unit Sales | Output | Month-Product-Region | SUM(MSPR Unit Sales) |
| 29 | MP Unit Sales | Output | Month-Product | SUM(MSP Unit Sales) |
| 30 | MP Sales Amount | Output | Month-Product | SUM(MSP Sales Amount) |
| 31 | Total Profit | Output | | SUM(Monthly Profit) |